# LATTICE-CELL : HYBRID APPROACH FOR TEXT CATEGORIZATION

Hichem Benfriha[1], Fatiha Barigou[2], Baghdad Atmani[3]

Laboratoire d'Informatique d'Oran, LIO, Oran University, Algeria
{[1]hichem.benfriha,[2]fatbarigou,[3]atmani.baghdad}@gmail.com

*ABSTRACT*

*In this paper, we propose a new text categorization framework based on Concepts Lattice and cellular automata. In this framework, concept structure are modeled by a Cellular Automaton for Symbolic Induction (CASI). Our objective is to reduce time categorization caused by the Concept Lattice. We examine, by experiments the performance of the proposed approach and compare it with other algorithms such as Naive Bayes and k nearest neighbors. The results show performance improvement while reducing time categorization.*

## 1. INTRODUCTION

Text Categorization is considered among the most important components in a research information system. In fact text Categorization allows the document organization in categories which accelerates and improves research and information retrieval[1]. Many different types of techniques have been used in text categorization, including probabilistic naive Bayesian methods, decision trees, k nearest neighbors and support vector machines. In this paper, the method we propose is based on concepts lattice and cellular automaton.

Concept lattice are useful to explore the voluminous information : The text/web classification [2], image retrieval [3], Information discovering and data warehouse classification [4]. The main advantage of the lattice concept is its completeness. But the counterpart, building and operating hierarchy are exponentially higher. So, the temporal and spatial complexity of concepts Lattice formulates the goal of this research. We provide a novel contribution to reduce this complexity by modeling the concepts lattice structure using cellular automaton CASI. We called our system LATTICE-CELL [5].

This paper is structured as follows. In Section 2, we recall basic results on concept lattice [6] and formal concept analysis (FCA) [7,8]. A detailed study of our LATTICE-CELL is presented in section 3. In this section we describe the architecture and the categorization methodology of the proposed system. Experimental results and comparative study are presented in sections 4 and 5 before concluding this work in Section 6.

## 2. BACKGROUND ON CONCEPTS LATTICES

The contents of Formal Concept Analysis (FCA) [9] are formal context, formal concept and the relation between the formal concepts. Formal context is defined as a triple K= (O, A, I), where O is the set of objects, A is the set of attributes and I the binary relationship between O and A, that is $I \subseteq O \times A$, oIa . In the formal context K, two mapping function X and Y is defined as follow :

$$A' = \{a \in A \mid oRa, \forall\, o \in A\} \qquad (1)$$

$$B' = \{o \in O \mid oRa, \forall\, a \in B\} \qquad (2)$$

They are called the Galois connection between O and A. If tuple (A, B) from P(G)×P(M) satisfied two conditions: A= X(B) and B=Y(A), or A=B' and B=A', (A, B) is called an concept from formal context K, denoted C = (A, B), B and A are called the Intension and Extension of concept C. Assume that C1 = (A1,B1) and C2 = (A2, B2) are two concepts, the order relation "≤" is defined as $C_1 \leq C_2 \Leftrightarrow B_2 \leq B_1$. C1 is the sub concept of $C_2$, $C_2$ is super concept of $C_1$. All concepts and their relations consist of a concept lattice L.

For example, figure 1 on the left shows the binary relation K= (O,A, I) (or formal context) and on the right the Hasse diagram of the concepts lattice derived from K. In the lattice, the concept $C_3$ = {(2, 6), (d)} in its Extension contains the objects 2 and 6, which have the property "d" in its Intension.

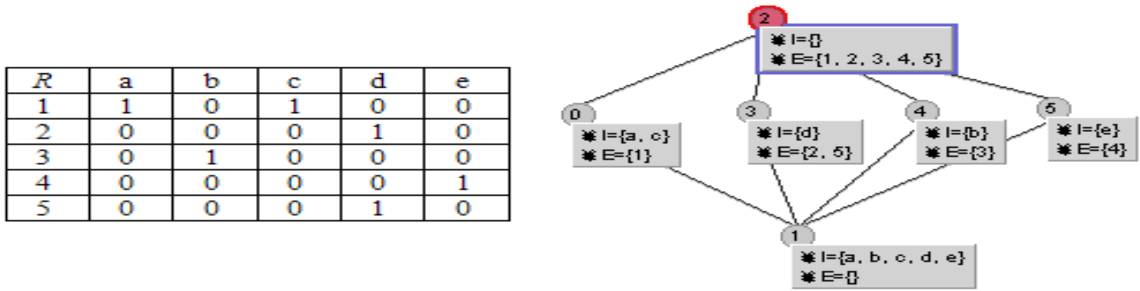

Figure 1. Left: Binary table K= (O = {1, 2, , . . . ,5}, A = {a,b, . . . , e}, I). Right: The Hasse diagram of the Concepts Lattice derived from K.

A variety of efficient algorithms exist for constructing the concept lattice from a binary table [9,10,11,12,13,14,15,16]. A classical distinction between them is made on two axes. The first one focuses on the way the initial data is acquired, it divides the methods on batch and incremental ones. The second axis distinguishes algorithms that construct the entire lattice, i.e., set of concepts and order (precedence) relation, from algorithms that only focus on concept set.

Authors in [15] proposed a novel paradigm for lattice construction that may be seen as a generalization of the incremental one. In fact, the suggested methods construct the final result in a divide-and-conquer mode, by splitting recursively the initial context B in two part B1 and B2 and then merging the lattices resulting (L1 and L2) from the processing of sub-contexts at various levels.

In this paper, we follow this approach to lattice algorithmics. The algorithm represents a complement of the well known algorithms of Noris, Chein and Ganter that compute the set of closed sets of a binary relation. When it is jointly applied with a concept computing algorithm, is an alternative to the batch procedures of Bordat and Lindig as well as to the incremental procedure of Godin, which constructs both the ground set and the Hasse diagram.

Our goal is to build a hierarchy of concepts, to use it for text categorization. Algorithm 1 presented in figure 2 shows the assembly phase when L1=B1≤1 and L2 = B2 ≤2 represents the two lattices corresponding to each part of the formal context and L = B ≤ represents the global Concepts Lattice.

```
Input:  L1= B1 ≤₁ , L2 = B2 ≤₂    /* Couple of lattices */
Output: L =B ≤                    /* The lattice of apposition context */
B := Ø
SORT(B1);SORT(B2)     /* sort of concept sets in an ascending order */
For each couple (Cᵢ, cj) in B1 x B2
    E := Ext(Cᵢ) ∩ Ext (Cⱼ)      /* computation of R */
    I := Int(Cᵢ) ∪ Int (Cⱼ)
    C := Find_Psi(E, Cᵢ, Cⱼ)  /* tentative retrieval of ƒ based on R */
    If C = NULL then
            C := Make_Concept (E, Int(Cᵢ),Ext(Cⱼ))
            B := B ∪ {C}

Find_LowerCovers (C, Cᵢ, Cⱼ)    /* detection of predecessors in L */
```

Figure 2. Algorithm 1 : Build the Global Lattice

## 3. LATTICE-CELL ARCHITECTURE

Our system called LATTICE-CELL consists of two modules as shown in figure 3. The first, is the preprocessing module that produces an index of words which is the vector representation or the formal concept. The outputs of the first module are the input of the second module which is going to generate the categorization model based on the Boolean representation for the lattice (cellular lattice).

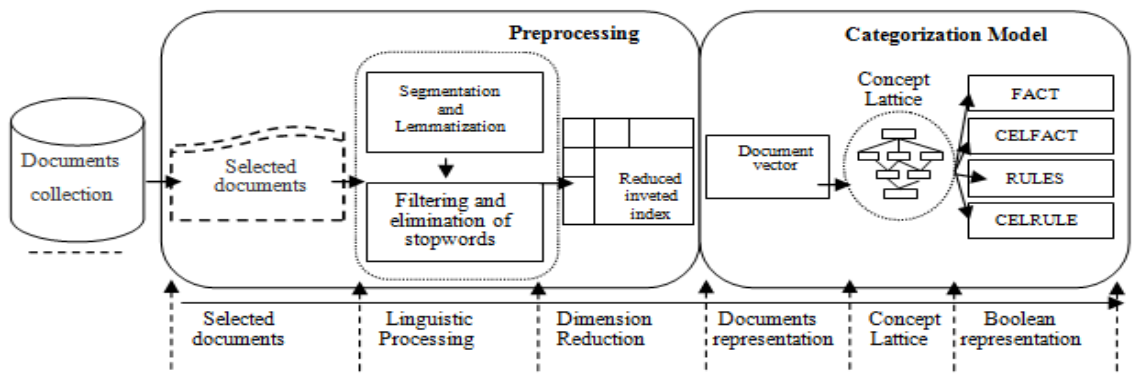

Figure 3. LATTICE-CELL Architecture

### 3.1. Preprocessing

The learners cannot operate on the documents as they are; the documents must be given internal representations that the learners can make sense of, once the learners build them. It is thus customary to transform all the documents (i.e., those used in the training phase, in the testing phase, or in the operational phase of the classifier) into internal representations. The step of preprocessing build a representation of the selected documents, all through a process of indexing is to extract terms representing the corpora [1].

**a) Linguistic processing** : This linguistic processing will be performed in 2 steps:

- **Segmentation and lemmatization** : This step is done by Tree Tagger[1] which will split the account into a set of units (words, numbers, punctuation, symbols) and reduce each word (name, verbs, adverbs, adjectives) into entities called first or canonical forms lemmas.

---
[1] http://www.ims.uni-Stuttgart.de/projekte/corplex/TreeTagger

- **Filtering and elimination of stopwords**: Cleaning is performed to remove the non-representative to retain only text content.

**b) Dimension Reduction :** This step can be performed by using a selection of words (lemmas) which keep only the words that are considered useful in the categorization according to some evaluation function, assigning a score to each word according to its discriminatory power , it suffices to keep only the words with the highest score and to significantly reduce the size of the space. In our case we chose to use the Information gain (IG) [17] as a measure of selection terms. For illustrative purposes we selected 9 documents that represent 3 categories (S: Sport, E: Economie and T: Television), and selected 6 words for index. Once the preprocessing step completed we obtain a vector representation with binary weighting for the 9 documents presented in table 1, which will represent the formal context.

Table 1. Vector representation with binary weighting

|       | Stade | Pays | Personnage | Ministre | Puissance | Visage |
|-------|-------|------|------------|----------|-----------|--------|
| Doc 1 | 1     | 1    | 0          | 0        | 0         | 0      |
| Doc 2 | 1     | 1    | 0          | 0        | 0         | 0      |
| Doc 3 | 0     | 0    | 0          | 0        | 0         | 1      |
| Doc 4 | 0     | 0    | 0          | 0        | 0         | 0      |
| Doc 5 | 0     | 0    | 0          | 1        | 1         | 0      |
| Doc 6 | 0     | 0    | 0          | 1        | 1         | 0      |
| Doc 7 | 1     | 0    | 0          | 0        | 0         | 1      |
| Doc 8 | 0     | 0    | 0          | 1        | 0         | 0      |
| Doc 9 | 0     | 0    | 1          | 0        | 0         | 0      |

**3.2. Boolean categorization model**

**3.2.1. Cellular Automaton**

CASI (Cellular Automaton for Symbolic Induction) [18,19] is a cellular Automaton that is made of two finite arbitrary long layers of finite state machines (cells) that are all identical. The operation of the system is synchronous, and the state of each cell at time t+1 depends only on the state of its vicinity cells, and on its own state at time t. This automaton, simulates the functioning of the basic cycle of an inference engine by using two finite layers of finite automata. The first layer, called CELFACT, is for representing the fact base, and the second layer, called CELRULE, is for representing the rule base. In each layer the content of a cell determines whether and how it participates in each inference step. At every step, a cell can be active or passive, and can take part in the inference or not. The states of cells are composed of two parts: EF and SF, and ER and SR, which are the input and output parts of the CELFACT cells, and of the CELRULE cells, respectively. Any cell i in the CELFACT layer with input EF (i) = 1 is regarded as representing an established fact. If EF (i) = 0, the represented fact has to be established. Any cell j of the CELRULE layer with input ER (j) = 0 is regarded as a candidate rule. When ER(j) = 1, the rule should not take part in the inference. Two incidence matrices called RE and RS define the neighborhood of cells. They represent the facts input relation respectively and the facts output relation. They are used in forward chaining.

The input relation, noted $iRE_j$, is formulated as follows: if (fact i ∈ Premise of rule j) then $iRE_j$ =1 else $iRE_j$ = 0.

The output relation, noted $iRS_j$, is formulated as follows: if (fact i ∈ Conclusion of rule j) then $iRS_j$ =1 else $iRS_j$ =0.The cellular automaton as a cycle of an inference engine made up of two local transitions functions δfact and δrule, where δfact corresponds to the evaluation, selection, and filtering phases and δrule corresponds to the execution phase.

- transitions functions δfact : $(EF,IF,SF,ER,IR,SR) \xrightarrow{\delta fact} (EF,IF,EF,ER+RE^T.EF,IR,SR)$

- transitions functions δrule: $(EF,IF,SF,ER,IR,SR) \xrightarrow{\delta rule} (EF+(R_S.ER),IF,SF,ER,IR,\overline{ER})$

### 3.2.2. Construction of the cellular Lattice

The vector presented in table 1 is used as input for the construction of cellular lattice. For this, we tuned algorithm 1 [15] by adding instructions to generate the cellular model representing the lattice. Algorithm 2 presented in figure 5 is the modified version of algorithm 1 to construct the cellular lattice.

```
Input: L1= B1 ≤₁ , L2 = B2 ≤₂         /* Couple of lattices */
Output: L =B ≤                        /* The lattice of apposition context */
B := Ø
R_c =1 ;            /* Initialization of the variable R_c corresponding to
                       the number of each rules created for each concept */

TCell_F(EF,IF,SF)[Ø]←(0,0,0); /* creation of the first layer CELFACT*/

TCell_R(ER,IR,SR)[Ø]←(0,0,0)  /*creation of the second layer CELRULE */

TCell_RE(RE)[Ø,0]←0;    /* creation for input incidence matrices R_E */

TCell_RS(RS)[Ø,0]←0;/* creation for the output incidence matrices
                                                                    R_S */

SORT(B1);SORT(B2)     /* sort of concept sets in an ascending order */
For each couple (C_i, C_j) in B1 x B2
    E := Ext(C_i) ∩ Ext (C_j);              /* computation of R */
    I := Int(C_i) U Int (C_j);
    C := Find_Psi(E, C_i, C_j)  /* tentative retrieval of f based on R */
    If C = NULL then

            C := Make_Concept (E, Int(C_i),Ext(C_j));
            TCell_F ((EF, IF, SF) [I_k])← (0,1,0);
            TCell_F ((EF, IF, SF) [E_k])← (0,1,0);

            TCell_R ((ER, IR, SR) [R_c])← (0,1,1);
            TCell_RE ([I_k, R_c]) ← 1;

            TCell_RS ([E_k, R_c]) ← 1;

            R_c= R_c +1;
B := B U {C};
Find_LowerCovers (C, C_i, C_j)    /* detection of predecessors in L */
```

Figure 5. Algorithm 2 : Construction of Cellular lattice

To construct cellular lattice, algorithm 2 uses for this purpose four functions :

1) **TCell _F** : for eatch intention $I_k$ and extension $E_k$, TCell_F creates two cells in the CELFACT layer: (EF, IF, SF) $[I_k]$ ← (0, 0, 0) to represent the fact $I_k$ ,

and (EF, IF, SF) [$E_k$] ← (0, 0, 0) to represent the fact $E_k$, (For the part $E_k$, TCell_F calculates the majority class of each extension), and where the concept has generated an empty word is in its intention or extension, TCell_F does not consider this concept.

2) **TCell _R** : for each concept $C_k$ not empty, TCell_R creates a cell in the CELRULE layer : (ER, IR, SR) ← (0, 1, 1) to represent the rule $R_C$.

3) **TCell_RE** : initialize the incidence matrice of input : RE [$I_k$, $R_C$] ← [1].

4) **TCell_RS** : initialize the incidence matrice of output: RS ← [$E_k$, $R_C$] ← [1].

By applying algorithm 2 for LATTICE-CELL system on the same representation of table 1, we obtain a cellular lattice represented by two layers of cellular automata CELFACT and CELRULE and incidences matrices RE and RS shown in figure 6.

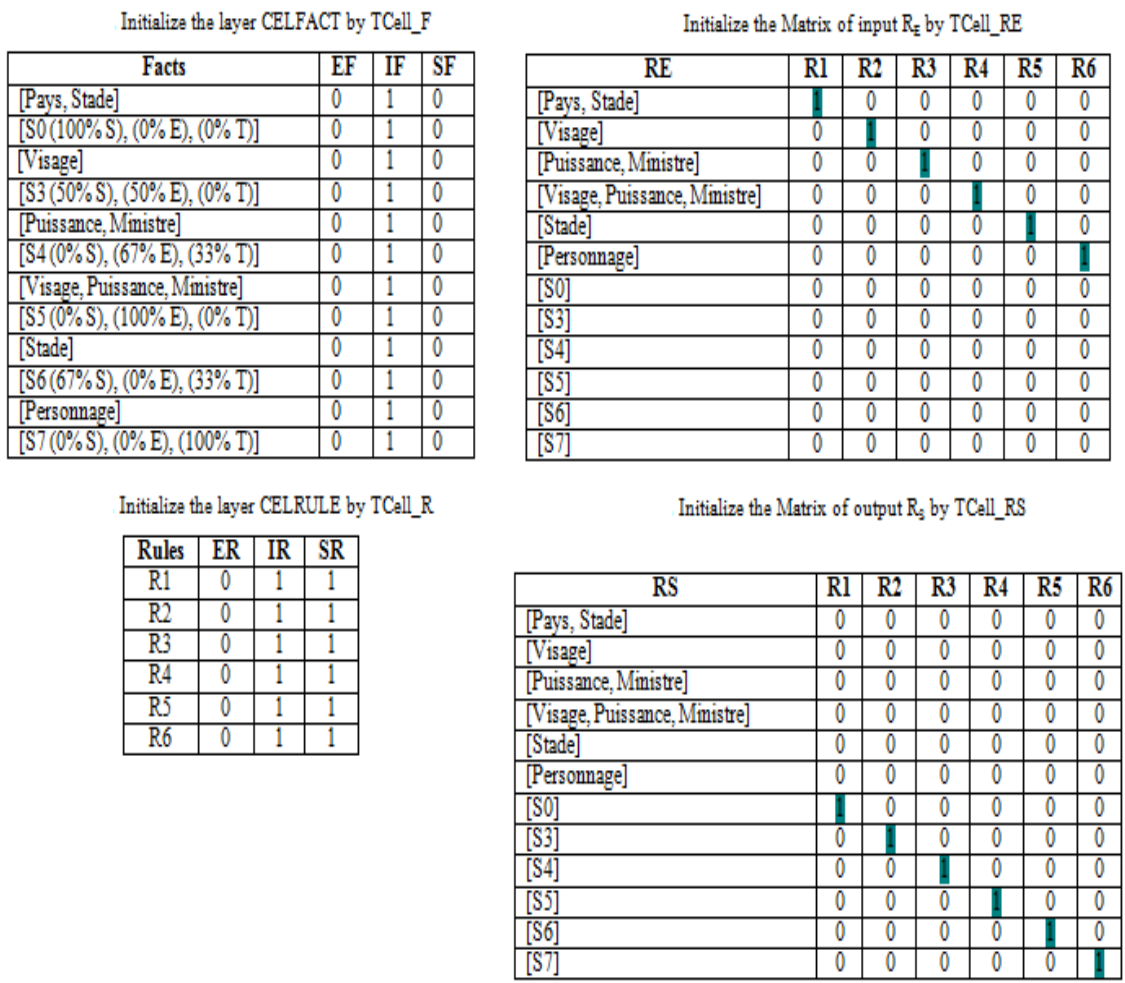

Figure 6. Cellular Lattice

## 4. EXPERIMENTATION AND DISCUSSION

LATTICE-CELL system has been tested on Corpora about (sport, economy and television)[2]. Tables 2 summarizes the documents distribution .

---
[2] http://deft08.limsi.fr/corpus-desc.php

Table 2. documents Distribution

|  | Sport | Economie | Television | Sample |
|---|---|---|---|---|
| Learning | 300 | 300 | 300 | **900** |
| Categorization | 154 | 154 | 154 | **462** |

### 4.1. Experiments results

Once the cellular lattice built, we proceed to the categorization phase which automatically determine documents category. To do this, we illustrate this step with an example of a document labeled "Economie", by following these steps: preprocessing, computing the similarity or initialization (Search similar documents), inference engine and majority voting to determine the document class.

### 4.1.1. Preprocessing

The categorization set will undergo the same preprocessing as learning set. We obtained the following vector.

Table 3. Vector of terms

| Stade | Pays | Personnage | Ministre | Puissance | Visage |
|---|---|---|---|---|---|
| 0 | 0 | 0 | 1 | 1 | 0 |

### 4.1.2. Computing the similarity (Initialization)

Document has been represented in vector form as points in a n-dimensional space, we determine the most similar vectors by calculating the distance between these points. There are different measures to calculate similarity as Jaccard, Cosine, Dice and Inner. Once one of these similarity measures used to calculate the distance between the vector of words in the table 3 and the word vectors in the layer CELFACT of figure 6. EF cells most similar vectors are initialized to 1, as shown in Table 4 and with the use of the similarity measure Inner:

Table 4. Initialization of the layer CELFACT.

| Facts | EF | IF | SF |
|---|---|---|---|
| [Pays, Stade] | 0 | 1 | 0 |
| [S0 (100% S), (0% E), (0% T)] | 0 | 1 | 0 |
| [Visage] | 0 | 1 | 0 |
| [S3 (50% S), (50% E), (0% T)] | 0 | 1 | 0 |
| [Puissance, Ministre] | **1** | 1 | 0 |
| [S4 (0% S), (67% E), (33% T)] | 0 | 1 | 0 |
| [Visage, Puissance, Ministre] | **1** | 1 | 0 |
| [S5 (0% S), (100% E), (0% T)] | 0 | 1 | 0 |
| [Stade] | 0 | 1 | 0 |
| [S6 (67% S), (0% E), (33% T)] | 0 | 1 | 0 |
| [Personnage] | 0 | 1 | 0 |
| [S7 (0% S), (0% E), (100% T)] | 0 | 1 | 0 |

### 4.1.3. Inference engine

Once the cells of the EF layer CELFACT instantiated to 1, the inference engine is initiated using the two transition functions δrule and δfact in forward chaining. After the execution of the transition function δrule, R4 and R5 are triggered as shown in the table 5 represents the layer CELRULE.

Table 5. The layer CELRULE after executing of δrule transition function

| Rules | ER | IR | SR |
|-------|----|----|----|
| R1    | 0  | 1  | 1  |
| R2    | 0  | 1  | 1  |
| R3    | 0  | 1  | 1  |
| R4    | **1** | 1 | 1 |
| R5    | **1** | 1 | 1 |
| R6    | 0  | 1  | 1  |

Once the execution of δrule transition function, the inference engine using the second function δfact it used for the execution phase. We note in table 6 that the EF cells peaks S4 and S5 are instantiated to 1.

Table 6. The layer CELFACT after the inference engine.

| Facts | EF | IF | SF |
|-------|----|----|----|
| [Pays, Stade] | 0 | 1 | 0 |
| [S0 (100% S), (0% E), (0% T)] | 0 | 1 | 0 |
| [Visage] | 0 | 1 | 0 |
| [S3 (50% S), (50% E), (0% T)] | 0 | 1 | 0 |
| [Puissance, Ministre] | 0 | 1 | 0 |
| [S4 (0% S), (67% E), (33% T)] | **1** | 1 | 0 |
| [Visage, Puissance, Ministre] | 0 | 1 | 0 |
| [S5 (0% S), (100% E), (0% T)] | **1** | 1 | 0 |
| [Stade] | 0 | 1 | 0 |
| [S6 (67% S), (0% E), (33% T)] | 0 | 1 | 0 |
| [Personnage] | 0 | 1 | 0 |
| [S7 (0% S), (0% E), (100% T)] | 0 | 1 | 0 |

### 4.1.4. Majority Vote

After the inference engine, a majority vote is launched to calculate the class membership of the document. If we take our example the two vertices are triggered: [S4 (0% S) (67% E) (33% T)] and [S5 (0% S), (100% E) (0% T )]. The system calculates the percentage of each categories and return to the user the highest 83.5% E (Economie) : for Sport : ((0 +0) / 2) = 0 %, for Economie: ((67 +100) / 2) = 83.5 % and for Television: ((33 +0) / 2) = 16.5 %.

### 4.2. Comparative experiments

To study the effect of the similarity measure of the performance of LATTICE-CELL, we tested four scenarios. In these scenarios, we varied the similarity measure, using Jaccard, Cosine, Dice and Inner and use GI as a function of reduction equal to 500 words. Table 7 presents the results of these experiments.

Table 7. Impact of similarity measures

|         | Precision | Recall | accuracy | Error | F-Mesure |
|---------|-----------|--------|----------|-------|----------|
| Jaccard | 0,34      | 0,39   | 0,54     | 0,45  | 0,36     |
| Cosinus | 0,39      | 0,33   | 0,60     | 0,40  | 0,36     |
| Inner   | 0,33      | 0,39   | 0,54     | 0,45  | 0,36     |
| Dice    | 0,32      | 0,69   | 0,42     | 0,57  | 0,44     |

We note, in Table 7 that the use of the Cosine measure gave a good performance in terms of accuracy and error rate. Table 8 shows the comparison between the LATTICE-CELL using the Cosine measure and other algorithms in the WEKA platform[3].

Table 8. Comparison of results

|              | Precision | Recall | accuracy | Error | F-Mesure |
|--------------|-----------|--------|----------|-------|----------|
| NaiiveBays   | 0,41      | 0,42   | 0,40     | 0,60  | 0,38     |
| ID3          | 0,38      | 0,36   | 0,36     | 0,64  | 0,35     |
| K-PPV        | 0,38      | 0,35   | 0,36     | 0,64  | 0,29     |
| SVM          | 0,38      | 0,37   | 0,38     | 0,62  | 0,36     |
| **LATTICE-CELL** | 0,39  | 0,33   | 0,60     | 0,40  | 0,36     |

LATTICE-CELL system allows us to obtain good results compared to other algorithms in terms of classification accuracy and error rate. We note that the system LATTICE-CELL improves nearly 7% K-PPV results in terms of F-measure and the proportion correctly classified documents (Acc) is higher with a 60% success rate, and the proportion of misfiled documents (Err) is significantly lower than the other algorithms as 40% error rate.

## 6. CONCLUSION

In this paper, we presented the functional architecture of our system LATTICE-CELL and the experimental study. LATTICE-CELL system can automatically acquire, represent and process knowledge extracted from sample texts in boolean form. The results show that the system LATTICE -CELL gives good results compared to other algorithms of  text categorizations such ID3, K-nearest neighbor, Naïve Bayes and support vector machines (SVM) in terms of accuracy and a much lower error rate. We managed to reduce the time categorization generated by the Concepts Lattice, with the new representation. The advantages of this modeling based on cellular automaton can be summarized as follows :

- The transition functions are easy to use, of low complexity, effective and robust regarding extreme values. Moreover, they are well adapted to situations with many concepts.

- The cellular model amounts to a simple set of transition functions and production rules, which not only make it possible to describe the problem at hand but also to build a classification function for class prediction.

- The incidences matrices RE and RS, facilitates the rules transformation into Boolean equivalent expressions and makes it possible, thereafter, to rely on elementary boolean algebra to test other simplifications.

---

[3] http://switch.dl.sourceforge.net/project/weka/weka-3-6-windows-jre/3.6.4/weka-3-6-4jre.exe